\documentstyle[prd,aps,preprint,amssymb,epsfig]{revtex}

\begin{document}


\def\Bbar{{\bar B}}
\def\Dbar{{\bar D}}
\def\qbar{{\bar q}}

\def\chihat{{\hat\chi}}

\def\calH{{\cal H}}
\def\calL{{\cal L}}
\def\calM{{\cal M}}
\def\calN{{\cal N}}

\def\zAn{z_A^{(n)}}
\def\xAn{x_A^{(n)}}
\def\MAn{M_A^{(n)}}
\def\NAn{N_A^{(n)}}
\def\aAn{\alpha_A^{(n)}}

\def\zn{z^{(n)}}
\def\xn{x^{(n)}}
\def\Mn{M^{(n)}}
\def\Nn{N^{(n)}}
\def\bn{\beta^{(n)}}
\def\chin{\chi^{(n)}}
\def\taun{\tau^{(n)}}

\def\dslash{{\partial\hspace{-2.5mm}/}}
\def\Wslash{{W\hspace{-3.5mm}/}}

\def\etal{{\it et al.}}
\def\ibid#1#2#3{{\it ibid.} {\bf #1} (#2) #3}

\def\epjc#1#2#3{Eur. Phys. J. C {\bf #1} (#2) #3}
\def\ijmpa#1#2#3{Int. J. Mod. Phys. A {\bf #1} (#2) #3}
\def\jhep#1#2#3{J. High Energy Phys. {\bf #1} (#2) #3}
\def\mpl#1#2#3{Mod. Phys. Lett. A {\bf #1} (#2) #3}
\def\npb#1#2#3{Nucl. Phys. {\bf B#1} (#2) #3}
\def\plb#1#2#3{Phys. Lett. B {\bf #1} (#2) #3}
\def\prd#1#2#3{Phys. Rev. D {\bf #1} (#2) #3}
\def\prl#1#2#3{Phys. Rev. Lett. {\bf #1} (#2) #3}
\def\rep#1#2#3{Phys. Rep. {\bf #1} (#2) #3}
\def\zpc#1#2#3{Z. Phys. {\bf #1} (#2) #3}


\title{$B\Bbar$ mixing with the bulk fields in the Randall-Sundrum model}
\author{Jong-Phil Lee\footnote{e-mail: jplee@phya.yonsei.ac.kr}}
\address{Department of Physics and IPAP, Yonsei University, Seoul, 120-749, Korea}

\tighten
\maketitle

\begin{abstract}

We calculate the $B\Bbar$ mixing in the Randall-Sundrum bulk model. 
In this model, all the Standard Model fields except the Higgs can reside in 
the bulk. 
Two suggestive models of "mixed" and "relaxed" scenarios are considered.
We find that the enhancement of the loop function is 0.51\% for the "relaxed" 
and 1.07\% for the "mixed" scenario when the first 4th KK modes are included,
for a bulk fermion mass parameter $\nu=-0.3$.

\end{abstract}
\pacs{}
\pagebreak

\section{Introduction}

The idea of extra dimension provides a very elegant explanation for the gauge
hierarchy problem of the standard model (SM).
Arkani-Hamed, Dimopoulos, and Dvali (ADD) suggested that there exist $n$ flat
extra dimensions with factorizable geometry \cite{ADD}.
In this model, the enormous Planck scale $M_{\rm Pl}$ is induced by the 
largeness of the extra dimensional volume $V_n$, via
$M_{Pl}^2=M_0^{n+2}V_n$.
New fundamental scale $M_0$ in $4+n$ dimension can be set down around 1 TeV, 
which resolves the hierarchy problem.
An alternative to ADD was soon proposed by Randall and Sundrum (RS) \cite{RS}
where the gauge hierarchy is explained by an exponential warp factor from a
5-dimensional non-factorizable geometry.
In this model, two branes are embedded at the boundaries of the ${\rm AdS}_5$ 
slice with a single $S_1/Z_2$ orbifold extra dimension.
\par
In the original model of ADD or RS, only the gravity can propagate in the bulk.
Effects of the bulk graviton appear in 4D world as an infinite tower of 
Kaluza-Klein (KK) modes.
Their phenomenological signatures at colliders are widely studied.
\par
A natural extension at the next step is to put the SM fields into the bulk.
In the universal extra dimension (UED) model by Appelquist, Cheng, and 
Dobrescu (ACD), all the SM fields are allowed to live in the extra dimensions
\cite{ACD}.
In the RS model, Goldberger and Wise tried to put the scalar fields into the
bulk, which has been developed to the bulk-stabilizing modulus field, or radion
\cite{GW}. 
They provided an open possibility that the SM matter fields might reside in the
extra dimension.
Later the gauge bosons are placed in the bulk in \cite{Pomarol}.
But the electroweak precision data constrain the first KK state of the gauge 
boson so strongly that the typical scale on the TeV brane goes up to about
100 TeV.
Apparently, this is not a good solution to the gauge hierarchy problem.
The problem was alleviated by putting fermion fields in the bulk while 
confining the Higgs to the TeV brane \cite{Chang}.
An attempt of constructing the bulk fermion fields was already done in 
\cite{Grossman} to explain the neutrino masses and mixings.
\par
Equipped with the RS-bulk SM fields, lots of works have been done to refine the
model and accommodate it to the existing data.
Here the fermion mass parameter $\nu\equiv m_\psi/k$ plays an important role,
where $m_\psi$ is the fermion bulk mass and ${\cal R}_5=-20k^2$ is the 
5-dimensional curvature \cite{Gherghetta,DHR}.
When placing the fermion fields in the bulk, there might occur large 
contributions to the flavor changing neutral current (FCNC) or the SM $\rho$ 
parameter.
Simple assumptions of a universal bulk fermion mass and the minimal flavor
violation where the CKM paradigm governs the flavor mixing can be a solution 
to this potential problems.
\par
However, when the Yukawa interaction is taken into account, non-negligible
mixing in the top sector can give a large contribution to the $\rho$ parameter.
Even worse is that the shifted mass spectrum in the top quark KK modes does not
guarantee the Glashow-Iliopoulos-Maiani (GIM) cancellation, which will cause
a disastrous FCNC.
Toward a more realistic model,
Hewett-Petriello-Rizzo (HPR) proposed the so called "mixed scenario" where the
first two generations of fermions are placed in the bulk while the third is
localized on the TeV brane \cite{HPR}.
While the HPR model reproduces the quark mass hierarchies $m_c/m_t$ and
$m_s/m_b$ without spoiling the $\rho$ parameter constraints, it seems unnatural
to confine only one family into the wall; there still remains a potential
danger of FCNC.
\par
As an alternative, a "relaxed" model was suggested by Kim-Kim-Song (KKS) where
the assumption of universal bulk fermion mass is slightly modified, retaining all
the fermions in the bulk \cite{KKS}.
In this approach, the $SU(2)$-singlet bottom quark field has a different bulk
mass $m_{\psi}'$.
They showed that introducing another parameter can be accommodated well to the 
electroweak precision data of $\Delta\rho\equiv\rho-\rho_{\rm SM}$ and 
$b\to s\gamma$.
\par
In this paper, we analyze $B\Bbar$ mixing based on the KKS.
With the advent of the $B$-factory era, we are now entering the age of precision
tests of flavor physics. 
Current world average of $\Delta M_B$, which parameterizes the $B\Bbar$ mixing,
is $\Delta M_B=0.502\pm0.006~{\rm ps}^{-1}$ \cite{HFAG} where the experimental
error is very small.
Relevant box diagrams are already studied in ACD-UED model, resulting in a 17\%
enhancement of the loop function \cite{Chakraverty,Buras}.
Present work will be a good comparison to the ACD-UED result.
It is also interesting to compare the results from KKS and HPR each other.
If the fermion mass parameter $\nu=-0.3$ is chosen (which is phenomenologically
viable in the literature), especially, mixing in the
top sector is rather small; the GIM cancellation is incomplete but we expect 
some remnant.
And the diagonalization of the mass matrix can be done perturbatively.
Naive thoughts lead to the idea that HPR will produce larger effects than KKS,
yielding to what amount of the excess of FCNC from the HPR would appear.
\par
The paper is organized as follows.
The RS-bulk model is reviewed in the next Section. 
We simply omit the gravity in the set up.
In Sec.\ III, mixing in the top sector is considered. 
We follow the description of KKS \cite{KKS}, and do not consider possible 
gauge boson mixing for simplicity.
The box diagrams for the $B\Bbar$ mixing are calculated in Sec.\ IV.
Section V contains the results and discussions.
A summary is given in Sec.\ VI.

\section{Setup of the model}

In the RS model, one spatial dimension is compactified on a
$S_1/Z_2$ orbifold of radius $r_c$ with a nonfactorizable geometry
\begin{equation}
ds^2=G_{MN}dx^M dx^N=e^{-2\sigma(\phi)}\eta_{\mu\nu}dx^\mu
dx^\nu+r_c^2 d\phi^2~,
\end{equation}
where the four-dimensional metric tensor is 
$\eta_{\mu\nu}={\rm diag}(1,-1,-1,-1)$, and 
$\sigma(\phi)=kr_c|\phi|,~0\le|\phi|\le\pi$. 
The model parameter $k$ is related to the 5-dimensional curvature 
${\cal R}_5=-20k^2$.
\par
In the original RS model, only the graviton can propagate through the bulk with 
its KK modes. 
We do not consider the graviton KK modes here because they are irrelevant for
$B\Bbar$ mixing.
We follow the model setup of KKS \cite{KKS}.
\par
Nonabelian gauge fields $A_M^a(x,\phi)$ can reside in the bulk via the 5D action
\begin{equation}
S_A=-\frac{1}{4}\int d^5x\sqrt{-G}G^{MK}G^{NL}F^a_{KL}F^a_{MN}~,
\end{equation}
where 
$F_{MN}^a=\partial_M A_N^a-\partial_N A_M^a-g_5\epsilon^{abc}A_M^b A_N^c$
($a,b,c=1,2,3$).
Choosing the gauge of $A_4(x,\phi)=0$ and assuming the KK expansion of $A_\mu$
to be
\begin{equation}
A_\mu^a(x,\phi)=\sum_{n=0}^\infty
A_\mu^{a(n)}(x)\frac{\chi_A^{(n)}(\phi)}{\sqrt{r_c}}~,
\end{equation}
we have a 4-dimensional effective action of massive KK gauge bosons as
\begin{equation}
S_A=\int d^4
x\sum_{n=0}^\infty\Bigg[-\frac{1}{4}\eta^{\mu\kappa}\eta^{\nu\lambda}
F^{a(n)}_{\kappa\lambda}F^{a(n)}_{\mu\nu}
-\frac{1}{2}M_A^{(n)2}\eta^{\mu\nu}A^{a(n)}_\mu
A^{a(n)}_\nu\Bigg]~,
\end{equation}
if the extra dimensional component $\chi_A^{(n)}(\phi)$ is the Bessel function
\begin{equation}
\chi_A^{(n)}=\frac{e^{\sigma(\phi)}}{N_A^{(n)}}\Big[J_1(z_A^{(n)}(\phi))
+\alpha_A^{(n)}Y_1(z_A^{(n)}(\phi))\Big]~.
\end{equation}
Here
\begin{equation}
\zAn(\phi)=\frac{M_A^{(n)}}{k}e^{\sigma(\phi)}~,
\end{equation}
where $\MAn$ is the mass of the $n$th KK mode of the gauge boson.
The continuity of $d\chi_A^{(n)}/d\phi$ at $\phi=0$ and $\phi=\pm\pi$ 
determines the mass spectrum and coefficient $\aAn$,
\begin{mathletters}
\begin{eqnarray}
\aAn&=&-\frac{J_1(\MAn/k)+(\MAn/k)J_1'(\MAn/k)}
             {Y_1(\MAn/k)+(\MAn/k)Y_1'(\MAn/k)}~~~({\rm at}~\phi=0)~,\\
J_1(\xAn)&+&\xAn J_1'(\xAn)
 +\aAn\Big[Y_1(\xAn)+\xAn Y_1'(\xAn)\Big]=0~~~({\rm at}~\phi=\pm\pi)~,
\end{eqnarray}
\end{mathletters}
where $\xAn\equiv\zAn(\phi=\pi)=(\MAn/k)e^{kr_c\pi}$.
The normalization constant
\begin{equation}
\NAn=\Bigg(\frac{e^{kr_c\pi}}{\xAn\sqrt{kr_c}}\Bigg)\sqrt{
{\zAn}^2\Big[J_1(\zAn)+\aAn Y_1(\zAn)\Big]^2
  \Bigg|^{\zAn(\phi=\pi)}_{\zAn(\phi=0)}}~,
\end{equation}
is obtained by the orthonormality condition
\begin{equation}
\int_{-\pi}^\pi d\phi \chi_A^{(n)} \chi_A^{(m)}=\delta^{mn}~.
\end{equation}
\par
Now consider the bulk fermion with arbitrary Dirac bulk mass.
A Dirac fermion field $\Psi$ with bulk mass $m_\psi$ is described by the 
5-dimensional action \cite{Grossman}
\begin{equation}
S_F=\int d^4x\int d\phi \sqrt{-G}\Bigg\{E^A_{\underline A}\Bigg[
 \frac{i}{2}{\bar\Psi}\gamma^{\underline A}(D_A-{\overleftarrow D}_A)\Psi
 \Bigg]
 -m_\psi{\rm sgn}(\phi){\bar\Psi}\Psi\Bigg\}~,
\label{SF}
\end{equation}
where $D_A$ is the covariant derivative, 
$\gamma^{\underline A}=(\gamma^\mu,i\gamma_5)$, and the inverse vielbein
$E^A_{\underline A}={\rm diag}(e^\sigma,e^\sigma,e^\sigma,e^\sigma,1/r_c)$.
The underlined uppercase Roman indices describe objects in the tangent frame.
Possible spin connection $\omega_{{\underline B\underline C}A}$ is omitted 
because it vanishes when the Hermitian conjugate is included.
\par
Integration by parts leads to 
\begin{eqnarray}
S_F&=&\int d^4 x\int d\phi r_c\Bigg\{e^{-3\sigma}\Big(
   {\bar\Psi}_Li\dslash~\Psi_L
   +{\bar\Psi}_Ri\dslash~\Psi_R\Big)\nonumber\\
&& -\frac{1}{2r_c}\Big[
   {\bar\Psi}_L(e^{-4\sigma}\partial_\phi+\partial_\phi e^{-4\phi})\Psi_R
  -{\bar\Psi}_R(e^{-4\sigma}\partial_\phi+\partial_\phi e^{-4\phi})\Psi_L\Big]
\nonumber\\
&& -e^{-4\sigma}m_\psi{\rm sgn}(\phi)
   ({\bar\Psi}_L\Psi_R+{\bar\Psi}_R\Psi_L)\Bigg\}~,
\label{SF2}
\end{eqnarray}
where the periodic boundary conditions of 
$\Psi_{L,R}(x,\pi)=\Psi_{L,R}(x,-\pi)$ are imposed.
Expanding $\Psi$ as
\begin{equation}
\Psi_{L,R}(x,\phi)=\sum_{n=0}^\infty\psi_{L,R}^{(n)}
 \frac{e^{2\sigma(\phi)}}{\sqrt{r_c}}{\hat f}_{L,R}^{(n)}(\phi)~,
\end{equation}
and requiring that
\begin{equation}
\int_{-\pi}^\pi~d\phi e^{\sigma(\phi)}
 {\hat f}^{(m)*}_L(\phi){\hat f}^{(n)}_L(\phi)
=\int_{-\pi}^\pi~d\phi e^{\sigma(\phi)}
 {\hat f}^{(m)*}_R(\phi){\hat f}^{(n)}_R(\phi)=\delta^{mn}~,
\end{equation}
\begin{equation}
\Bigg(\pm\frac{1}{r_c}\partial_\phi-m_\psi\Bigg){\hat f}^{(n)}_{L,R}(\phi)
=-\Mn_f e^\sigma{\hat f}^{(n)}_{R,L}(\phi)~,
\end{equation}
we have the effective action for the massive Dirac fermions
\begin{equation}
S_F=\sum_{n=0}^\infty\int d^4x\Big[
 {\bar\psi}^{(n)}(x)i\dslash~\psi^{(n)}(x)-\Mn_f{\bar\psi}^{(n)}(x)\psi^{(n)}(x)
 \Big]~.
\end{equation}
The $Z_2$-symmetry of $S_F$ in (\ref{SF}) requires that
${\bar\Psi}\Psi={\bar\Psi}_L\Psi_R+{\bar\Psi}_R\Psi_L$ be $Z_2$-odd.
This requirement is satisfied if we impose the opposite $Z_2$-parity on
${\hat f}^{(n)}_{L,R}$.
We fix ${\hat f}^{(n)}_{L(R)}$ to be $Z_2$-even(odd).
Introducing $\nu\equiv m_\psi/k$ of order 1, the solutions are, for $n\ne 0$,
\begin{mathletters}
\begin{eqnarray}
{\hat f}^{(n)}_L&\equiv&\chin(\phi)=\frac{e^{\sigma/2}}{\Nn_\chi}\Big[
 J_{1/2-\nu}(\zn)+\bn_\chi Y_{1/2-\nu}(\zn)\Big]~,\\
{\hat f}^{(n)}_R&\equiv&\taun(\phi)=\frac{e^{\sigma/2}}{\Nn_\tau}\Big[
 J_{1/2+\nu}(\zn)+\bn_\tau Y_{1/2+\nu}(\zn)\Big]~,\\
\end{eqnarray}
\end{mathletters}
where $\zn_f(\phi)\equiv(\Mn_f/k)e^{\sigma(\phi)}$ and for $n=0$,
\begin{equation}
\chi^{(0)}(\phi)=\frac{e^{\nu\sigma(\phi)}}{N^{(0)}_\chi}~,~~~
\tau^{(0)}(\phi)=0~.
\end{equation}
Mass spectrum and the coefficients $\bn_{\chi,\tau}$ as well as the 
normalization constants $\Nn_{\chi,\tau}$ are determined by the boundary
conditions:
\begin{equation}
0=\Bigg(\frac{d}{d\phi}-m_\psi r_c\Bigg)\chin|_{\phi=0,\pi}
 =\taun|_{\phi=0,\pi}~.
\end{equation}
Explicitly,
\begin{equation}
\bn_\chi=-\frac{J_{-(1/2+\nu)}(\Mn_f/k)}{Y_{-(1/2+\nu)}(\Mn_f/k)}~,
\end{equation}
from the boundary conditions at $\phi=0$, and
\begin{equation}
J_{-(1/2+\nu)}(\xn_f)+\bn_\chi Y_{-(1/2+\nu)}(\xn_f)=0~.
\end{equation}
at $\phi=\pi$, where $\xn_f\equiv\zn_f(\phi=\pi)$.
Similar expressions can be given for the $\taun(\phi)$ sector, and
the left- and right-handed excitation masses are degenerate to $\Mn_f$ for a 
given $n$.
The normalization constants are
\begin{equation}
\Nn_{\chi,\tau}=\Bigg(\frac{e^{kr_c\pi}}{\xn_f\sqrt{kr_c}}\Bigg)\sqrt{
 {\zn_f}^2\Big[J_{1/2\mp\nu}(\zn_f)+\bn_{\chi,\tau} Y_{1/2\mp\nu}(\zn_f)\Big]^2
 \Bigg|^{\zn_f(\phi=\pi)}_{\zn_f(\phi=0)}~.
 }
\end{equation}
\par
Some difficulties arise when the fermions are included in the bulk because
the fermion field contents must be doubled.
In our setup of (\ref{SF2}), a fermion which belongs to a specific 
representation of a gauge group should be vector-like, possessing both left- 
and right-handed chiralities.
For each generation, we have an $SU(2)$-doublet $Q=(q_u,q_d)^T$ and two $SU(2)$-
singlets $u$ and $d$ where both chiralities are allowed for all of them.
Explicitly,
\begin{mathletters}
\begin{eqnarray}
Q(x,\phi)&=&Q_L+Q_R=\sum_n\frac{e^{2\sigma(\phi)}}{\sqrt{r_c}}\Big[
 Q^{(n)}_L(x)\chin(\phi)+Q^{(n)}_R(x)\taun(\phi)\Big]~,\\
u(x,\phi)&=&u_L+u_R=\sum_n\frac{e^{2\sigma(\phi)}}{\sqrt{r_c}}\Big[
 u^{(n)}_L(x)\taun(\phi)+u^{(n)}_R(x)\chin(\phi)\Big]~,
\end{eqnarray}
\end{mathletters}
where we have assigned $\chin$ to the left-handed $SU(2)$ doublet and the 
right-handed $SU(2)$ singlet in order to accommodate the SM fermion fields.
The $SU(2)$ singlet down part $d(x,\phi)$ has the same KK decomposition as
$u(x,\phi)$.
\par
With these bulk-SM fields, the charged current interactions look like
\begin{eqnarray}
S_{ffW}&=&\int d^5 x\sqrt{-G}e^\sigma\frac{g_5}{\sqrt{2}}\big[
          \qbar_u\Wslash^+ q_d +{\rm h.c.}\big]\nonumber\\
&=&\int d^4 x\frac{g}{\sqrt{2}}\sum_{l=0}^\infty\Bigg[\sum_{n,m=0}^\infty
 \qbar^{(n)}_{uL}\Wslash^{+(l)}q^{(m)}_{dL} C^{ffW}_{nml}\nonumber\\
&& +\sum_{n,m=1}^\infty\qbar^{(n)}_{uR}\Wslash^{+(l)}q^{(m)}_{dR}\Bigg(
   \sqrt{2\pi}\int_{-\pi}^\pi d\phi~e^\sigma\taun\tau^{(m)}\chi_A^{(l)}\Bigg)
\Bigg]+{\rm h.c.}~,
\end{eqnarray}
where $g=g_5/\sqrt{2\pi r_c}$ and the coupling constant $C^{ffW}_{nml}$ is
\begin{equation}
C^{ffW}_{nml}=\sqrt{2\pi}\int_{-\pi}^\pi d\phi~e^\sigma
 \chin(\phi)\chi^{(m)}(\phi)\chi_A^{(l)}(\phi)~.
\label{CffW}
\end{equation}

\section{Mixing in the Top sector}

For the bulk fermions, there are two sources of physical fermion mass spectrum.
One is the KK mass which is proportional to $\sim\qbar_{uL}q_{uR}$ or
$\sim{\bar u}_Lu_R$.
The other is the Yukawa interaction which relates the $SU(2)$ doublet with the
singlet, $\sim\qbar_{uL}u_R$, as in the SM.
This difference results in mixing among the fermion KK modes.
If the quark masses were sufficiently smaller than the KK mass scale, 
this mixing effects might be negligible.
In this case, all the fermion KK mass spectrum would be almost degenerate 
assuming that the model is minimal with a single parameter $m_\psi$ for the 
universal bulk fermion mass.
The exact degeneracy between the KK masses for the $T_3=\pm1/2$ fermions is 
responsible for the vanishing contribution to $\Delta\rho$.
Furthermore, we can expect the GIM cancellation at each level of KK modes
because all the up- (and down-) type quarks are degenerate.
\par
The presence of top quark which is very heavy makes the mixing non-negligible.
Consequently, degenerate top and bottom KK masses are shifted.
They can cause a large $\Delta\rho$, because the quantum correction is 
proportional to the squared mass difference between up- and down-type quarks.
Adding higher KK modes makes the problem worse.
\par
As mentioned in the Introduction, HRP proposed the "mixed" scenario where the
3rd-generation fermions are confined to the TeV brane while the other two
can propagate in the bulk.
Instead, we adopt the KKS model of \cite{KKS} where the $SU(2)$-singlet bottom
quark field has a different bulk fermion mass $m_\psi'$.
The model is recapitulated in the 5D action for the 3rd-generation quarks
\begin{eqnarray}
S&=&\int d^4x\int d\phi\sqrt{-G}\Big[E^A_a(i{\bar Q}\gamma^a{\cal D}_A Q
+i{\bar t}\gamma^a{\cal D}_A t +i{\bar b}\gamma^a{\cal D}_A b)\nonumber\\
&&
-{\rm sgn}(\phi)(m_\psi\{{\bar Q}Q+{\bar t}t\}+m_\psi'{\bar b}b)\Big]~.
\end{eqnarray}
Expanding the fermion fields with the mode functions and integrating over $\phi$
yield the bulk fermion KK masses as 
\begin{equation}
{\cal L}=-\sum_{n=1}^\infty k_{EW}\Big[\xn_f(\nu)\Big(
 \qbar^{(n)}_{tL}q^{(n)}_{tR}+\qbar^{(n)}_{bL}q^{(n)}_{bR}
    +{\bar t}^{(n)}_Lt^{(n)}_R\Big)
 +\xn_f(\nu'){\bar b}^{(n)}_Lb^{(n)}_R\Big]+{\rm h.c.}~,
\label{KKmass}
\end{equation}
where $k_{EW}=k\epsilon\equiv k e^{-kr_c\pi}$, and $\nu'\equiv m_\psi'/k$.
\par
Now consider the Yukawa interactions associating the Higgs boson.
Ordinary Higgs mechanism works here so that the SM particles acquire masses.
It has been argued, however, that if there were bulk Higgs fields, the hierarchy
problem remains unsolved \cite{Chang} or the observed $W$ and $Z$ mass relations
cannot be reproduced \cite{DHR}.
We assume here that the Higgs field is confined to the TeV brane.
Then the 5D action for Yukawa interactions is
\begin{equation}
S_{ffH}=-\int d^5x\sqrt{-G}\Bigg[
 \frac{\lambda_5^b}{k}{\bar Q}(x,\phi)\cdot H(x)b(x,\phi)
 +\frac{\lambda_5^t}{k}\epsilon^{ab}{\bar Q}(x,\phi)_a\cdot H(x)_b t(x,\phi)
 +{\rm h.c.}\Bigg]\delta(\phi-\pi)~,
\end{equation}
where $\lambda^{b,t}_5$ is the 5D Yukawa coupling.
After the spontaneous symmetry breaking, the Higgs field shifts 
$H^0\to v_5+H'^0$, and the 4D effective Lagrangian becomes
\begin{eqnarray}
{\cal L}_{\rm eff}&=&\frac{\lambda_t v}{\sqrt{2}}\big(
 \qbar^{(0)}_{tL}+{\hat\chi}_1\qbar^{(1)}_{tL}+\cdots\big)\big(
 t^{(0)}_R+{\hat\chi}_1t^{(1)}_R+\cdots\big)\nonumber\\
&&
 +\frac{\lambda_b v}{\sqrt{2}}\big(
 \qbar^{(0)}_{bL}+{\hat\chi'}_1\qbar^{(1)}_{bL}+\cdots\big)\big(
 b^{(0)}_R+{\hat\chi'}_1b^{(1)}_R+\cdots\big)~,
\label{Yukawamass}
\end{eqnarray}
where $\lambda_{t,b}=\lambda_5^{t,b}(1+2\nu)/2(1-\epsilon^{1+2\nu})$,
$v=\epsilon v_5$, ${\hat\chi}_n\equiv\chin(\pi,\nu)/\chi^{(0)}(\pi,\nu)$, and
${\hat\chi'}_n$ are with $\nu'$.
\par
From Eqs.\ (\ref{KKmass}) and (\ref{Yukawamass}), the physical mass term for
the top sector is
\begin{equation}
\calL_{\rm mass}^{t}=-
\left(\begin{array}{ccc|cc}
{\bar t}^{(0)}_R & {\bar t}^{(1)}_R & \cdots~&~\qbar^{(1)}_{tR} & \cdots
\end{array}\right)
\calM_t
\left(\begin{array}{c}
q^{(0)}_{tL}\\q^{(1)}_{tL}\\\vdots\\ - \\ t^{(1)}_L\\\vdots
\end{array}\right)~.
\end{equation}
The bottom sector shows very similar mass terms.
Let the number of the KK states be $n_\infty$ which is, in principle, infinite.
Then the $(2n_\infty+1)\times(2n_\infty+1)$ matrix $\calM_t$ has the form of
\begin{equation}
\calM_t=
\left(\begin{array}{cc}
\calM^t_Y & \calM^t_{KK}\\
\calM^{q_t}_{KK} & 0 
\end{array}\right)~,
\label{Mt}
\end{equation}
where the $(n_\infty+1)\times(n_\infty+1)$ matrix $\calM^t_Y$ is from 
Yukawa mass term and the $(n_\infty+1)\times n_\infty$ matrix $\calM_{KK}$
from KK masses.
They are given by
\begin{mathletters}
\begin{eqnarray}
\calM^t_Y&=&m_{t,0}\left(\begin{array}{cccc}
1 & \chihat_1 & \chihat_2 & \cdots \\
\chihat_1 & \chihat_1^2 & \chihat_1\chihat_2 & \cdots \\
\chihat_2 & \chihat_2\chihat_1 & \chihat_2^2 & \cdots \\
\vdots & \vdots & \vdots & 
\end{array}\right)~,\\
\calM^{q_t}_{KK}&=&k_{EW}\left(\begin{array}{cccc}
0 & x^{(1)}_f & 0 & \cdots \\
0 & 0 & x^{(2)}_f & \cdots \\
0 & 0 & 0 & \cdots \\
\vdots & \vdots & \vdots & 
\end{array}\right)=\calM^t_{KK}~,
\end{eqnarray}
\end{mathletters}
where $m_{t,0}=\lambda_t v/\sqrt{2}$.
\par
Now the physical mass eigenstates $u^{\prime(n)}_L$ are obtained by 
diagonalizing (\ref{Mt}) through an orthogonal matrix $\calN$:
\begin{equation}
\left(\begin{array}{c}
u^{\prime(0)}_L \\ u^{\prime(1)}_L \\ u^{\prime(2)}_L \\ \vdots
\end{array}\right)
=\calN\left(\begin{array}{c}
q^{(0)}_{uL} \\ q^{(1)}_{uL} \\ \vdots \\ - \\ u^{(1)}_L \\ \vdots
\end{array}\right)~.
\end{equation}
For example, $q^{(n)}_{uL}$ can be written in terms of KK mass eigenstates 
$u^{\prime({\underline j})}_L$:
\begin{equation}
q^{(n)}_{uL}=\sum_{{\underline j}=0}^{2n_\infty}\calN_{({\underline j},n)}
u^{\prime({\underline j})}_L~.
\end{equation}
Hereafter, the underlined index runs from zero to $2n_\infty$.
In terms of mass eigenstates, the $Wtb$-vertex in 4D becomes
\begin{equation}
\calL=\frac{g}{\sqrt{2}}V_{tb}\left(\sum_{m=0}^{n_\infty}C^{btW}_{0ml}
\calN_{({\underline j},m)}\right){\bar b}^{(0)}_L\gamma^\mu 
 t^{\prime({\underline j})}_L W^{(l)}_\mu +{\rm h.c.}
\end{equation}
\par
For light quarks ($m_{q,0}=0$), $\calM_q$ can be analytically diagonalized
to $\calM_q\to{\rm diag}(m^{(0)}_q, M^{(0)}_1, M^{(0)}_2, \cdots)$ where
\begin{equation}
m^{(0)}_q=0~,~~~M^{(0)}_n=M^{(0)}_{n_\infty+n}=\xn_f k_{EW}~.
\end{equation}
The corresponding orthonormal matrix $\calN^{(0)}$ is
\begin{eqnarray}
\calN^{(0)}_{(0,0)}&=&1~,~~~
\calN^{(0)}_{(0,{\underline n})}=\calN^{(0)}_{({\underline n},0)}=0~,
\nonumber\\
\calN^{(0)}_{(n,m)}&=&-\calN^{(0)}_{(n_\infty+n,m)}
 =\frac{\delta_{nm}}{\sqrt{2}}~.
\end{eqnarray}
\par
On the other hand, the diagonalization of $\calM_t$ of the top sector is 
nontrivial since $m_{t,0}$ is heavy.
However, an approximate calculation is possible unless $\chihat_n^2$'s are 
much larger than unity, with a small parameter $m_{t,0}/k_{EW}$.
It is shown in \cite{KKS} that a perturbative diagonalization can be done for
$\nu\gtrsim -0.3$.
Up to the leading order of the expanding parameter 
$\delta\equiv m_{t,0}/k_{EW}$, the mass eigenvalues are
\begin{eqnarray}
m_q&=&m_0~,~~~ M_n=k_{EW}\Bigg(\xn_f+\frac{\chihat_n^2}{2}\delta\Bigg)~,
\nonumber\\
M_{n_\infty+n}&=&k_{EW}\Bigg(\xn_f-\frac{\chihat_n^2}{2}\delta\Bigg)~,
\end{eqnarray}
and the orthonormal matrix $\calN$ are
\begin{equation}
\calN_{({\underline n},{\underline m})}\equiv
\calN^{(0)}_{({\underline n},{\underline m})}+
\calN^{(1)}_{({\underline n},{\underline m})}\frac{\delta}{\sqrt{2}}~,
\label{Ntop1}
\end{equation}
where
\begin{eqnarray}
\calN^{(1)}_{(0,0)}&\simeq&0~,~~~\calN^{(1)}_{(0,n)}\simeq 0~,~~~
\calN^{(1)}_{(n,0)}\simeq\calN^{(1)}_{(n_\infty+n,0)}
 \simeq\frac{\chihat_n^2}{\xn_f}~,\nonumber\\
\calN^{(1)}_{(n,n)}&\simeq&\calN^{(1)}_{(n_\infty+n,n)}
 \simeq\frac{\chihat_n^2}{4\xn_f}~,\nonumber\\
\calN^{(1)}_{(n,m)}&\simeq&\calN^{(1)}_{(n_\infty+n,m)}
 \simeq\frac{\xn_f\chihat_n\chihat_m}{\sqrt{2}({\xn_f}^2-{x^{(m)}_f}^2)}
~~~({\rm for}~n\ne m)~.
\label{Ntop2}
\end{eqnarray}

\section{$B\Bbar$ mixing and the box diagrams}

The $B\Bbar$ mixing is parametrized by the physical mass difference
\begin{eqnarray}
\Delta M_B&=&\frac{1}{M_B}|\langle\Bbar^0|\calH_{\rm eff}(\Delta B=2)|B^0
\rangle| \nonumber\\
&=&\frac{G_F^2M_W^2}{6\pi^2}|V_{tb}^*V_{td}|^2
   {\hat B}_B f_B^2M_B\eta_B S_0(x_t)~.
\end{eqnarray}
where $M_B ~(M_W)$ is the $B~(W)$ mass, $\eta_B$ the QCD corrections, and
${\hat B}_B$ and $f_B$ are the bag parameter and decay constant, respectively.
The loop function $S_0(x_t)$ which recapitulates the box diagrams is
\begin{equation}
S_0(x_t)=\frac{4x_t-11x_t+x_t^3}{4(1-x_t)^2}
 -\frac{3x_t^3\ln x_t}{2(1-x_t)^3}~,
\end{equation}
in the SM, with $x_t\equiv (m_t/M_W)^2$.
\par
\begin{figure}
\vskip 2cm
\begin{center}
\epsfig{file=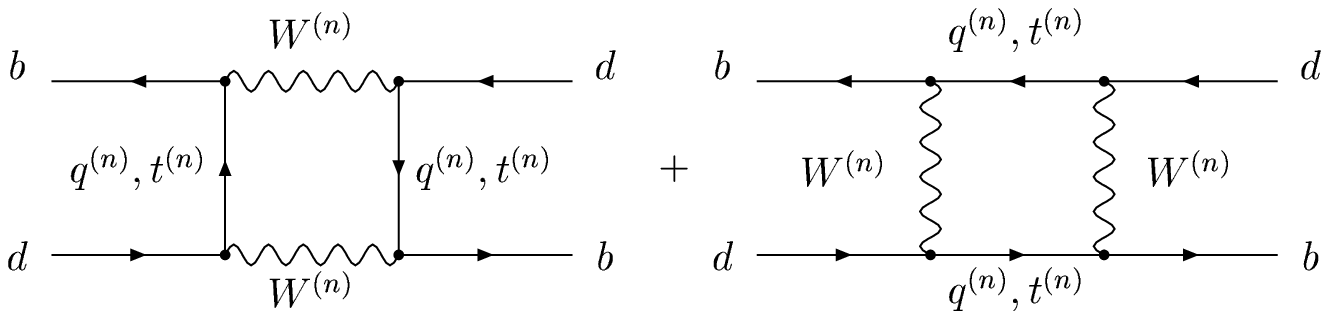}
\end{center}
\caption{}
\label{box}
\end{figure}

In the RS-bulk model, bulk fermions and gauge bosons contribute to the box
diagrams, as shown in Fig.\ \ref{box}.
Since the external fermions are the zero modes, relevant vertex factor is
$C^{ffW}_{m0n}$ multiplied by the mass-diagonalizing orthonormal matrix 
$\calN$.
\par
Introducing the effective couplings
\begin{mathletters}
\begin{eqnarray}
T^{ij}&\equiv&\Big(\calN\cdot C^{b(d)tW}\Big)^{ij}~,\\
Q^{ij}&\equiv&\Big(\calN\cdot C^{b(d)qW}\Big)^{ij}~,
\end{eqnarray}
\end{mathletters}
the loop function has the form of
\begin{eqnarray}
S(m,n;i,j)&=&M_W^2\sum\left[\left(
 1+\frac{m_{t^{(i)}}^2 m_{t^{(j)}}^2}{4{M^{(m)}_A}^2 {\Mn_A}^2}\right)
 T^{im}T^{in}T^{jn}T^{jm}I(W^{(m)}W^{(n)}t^{(i)}t^{(j)})\right.\nonumber\\
&&
 +\left(1+\frac{m_{q^{(i)}}^2 m_{q^{(j)}}^2}{4{M^{(m)}_A}^2 {\Mn_A}^2}\right)
 Q^{im}Q^{in}Q^{jn}Q^{jm}I(W^{(m)}W^{(n)}q^{(i)}q^{(j)})\nonumber\\
&&
-\left(2+\frac{m_{q^{(i)}}^2 m_{t^{(j)}}^2+m_{t^{(i)}}^2 m_{q^{(j)}}^2}
 {4{M^{(m)}_A}^2 {\Mn_A}^2}\right)Q^{im}Q^{in}T^{jn}T^{jm}
 I(W^{(m)}W^{(n)}q^{(i)}t^{(j)})\nonumber\\
&&
+\left(\frac{1}{{M_A^{(m)}}^2}+\frac{1}{{\Mn_A}^2}\right)\Big\{
 -m_{t^{(i)}}^2 m_{t^{(j)}}^2
   T^{im}T^{in}T^{jn}T^{jm}I(W^{(n)}\phi^{(m)}t^{(i)}t^{(j)})\nonumber\\
&&
 -m_{q^{(i)}}^2 m_{q^{(j)}}^2
   Q^{im}Q^{in}Q^{jn}Q^{jm}I(W^{(n)}\phi^{(m)}q^{(i)}q^{(j)})\nonumber\\
&&\left.
+(m_{q^{(i)}}^2 m_{t^{(j)}}^2+m_{t^{(i)}}^2 m_{q^{(j)}}^2)
 Q^{im}Q^{in}T^{jn}T^{jm}I(W^{(n)}\phi^{(m)}q^{(i)}t^{(j)})\Big\}\right]~.
\end{eqnarray}
Here we separate the longitudinal part of $W^{(n)}$, denoted by $\phi^{(n)}$,
for the calculational convenience, and
\begin{mathletters}
\begin{eqnarray}
I(W^{(m)}W^{(n)}q^{(i)}q^{(j)})&=&
 \int_0^1 dx dy\left[
 \frac{M_{mn}^2+m_{q^{(i)}q^{(j)}}^2}{(M_{mn}^2-m_{q^{(i)}q^{(j)}}^2)^2}
+\frac{2m_{q^{(i)}q^{(j)}}^2 M_{mn}^2}{(M_{mn}^2-m_{q^{(i)}q^{(j)}}^2)^3}
\ln\frac{m_{q^{(i)}q^{(j)}}^2}{M_{mn}^2}\right]~,\\
I(W^{(n)}\phi^{(m)}q^{(i)}q^{(j)})&=&
 \int_0^1 dx dy\left[
 \frac{2}{(M_{mn}^2-m_{q^{(i)}q^{(j)}}^2)^2}
 +\frac{M_{mn}^2+m_{q^{(i)}q^{(j)}}^2}{(M_{mn}^2-m_{q^{(i)}q^{(j)}}^2)^3}
\ln\frac{m_{q^{(i)}q^{(j)}}^2}{M_{mn}^2}\right]~,
\end{eqnarray}
\end{mathletters}
where
\begin{equation}
m_{q^{(i)}q^{(j)}}^2\equiv xm_{q^{(i)}}^2+(1-x)m_{q^{(j)}}^2~,~~~
M_{mn}^2\equiv y {M^{(m)}_A}^2+(1-y){\Mn_A}^2~.
\end{equation}

\section{Results and Discussions}

For the numerical results, we used $kr_c=11.5$, $k_{EW}=1~{\rm TeV}$, and
$\nu=-0.3$.
The choice of $\nu=-0.3$ allows us to use Eqs.\ (\ref{Ntop1}) and 
(\ref{Ntop2});
for $\nu\le-0.4$, only the numerical diagonalization of the mass matrix is 
reliable.
One question may arise whether this choice would spoil the $\Delta\rho$
accommodation.
This, however, would not be the case because we have another adjustable
parameter $\nu'$ for the $b$-sector in KKS model.
Even in the case of $(\nu,\nu')=(-0.3,-0.6)$ the mass differences between top
and bottom KK modes are much smaller than 1 TeV \cite{Fig4}.
We also give the results for $\nu=-0.2,~-0.1$ for comparisons.
\par
\begin{figure}
\vskip 2cm
\begin{center}
\begin{tabular}{cc}
\epsfig{file=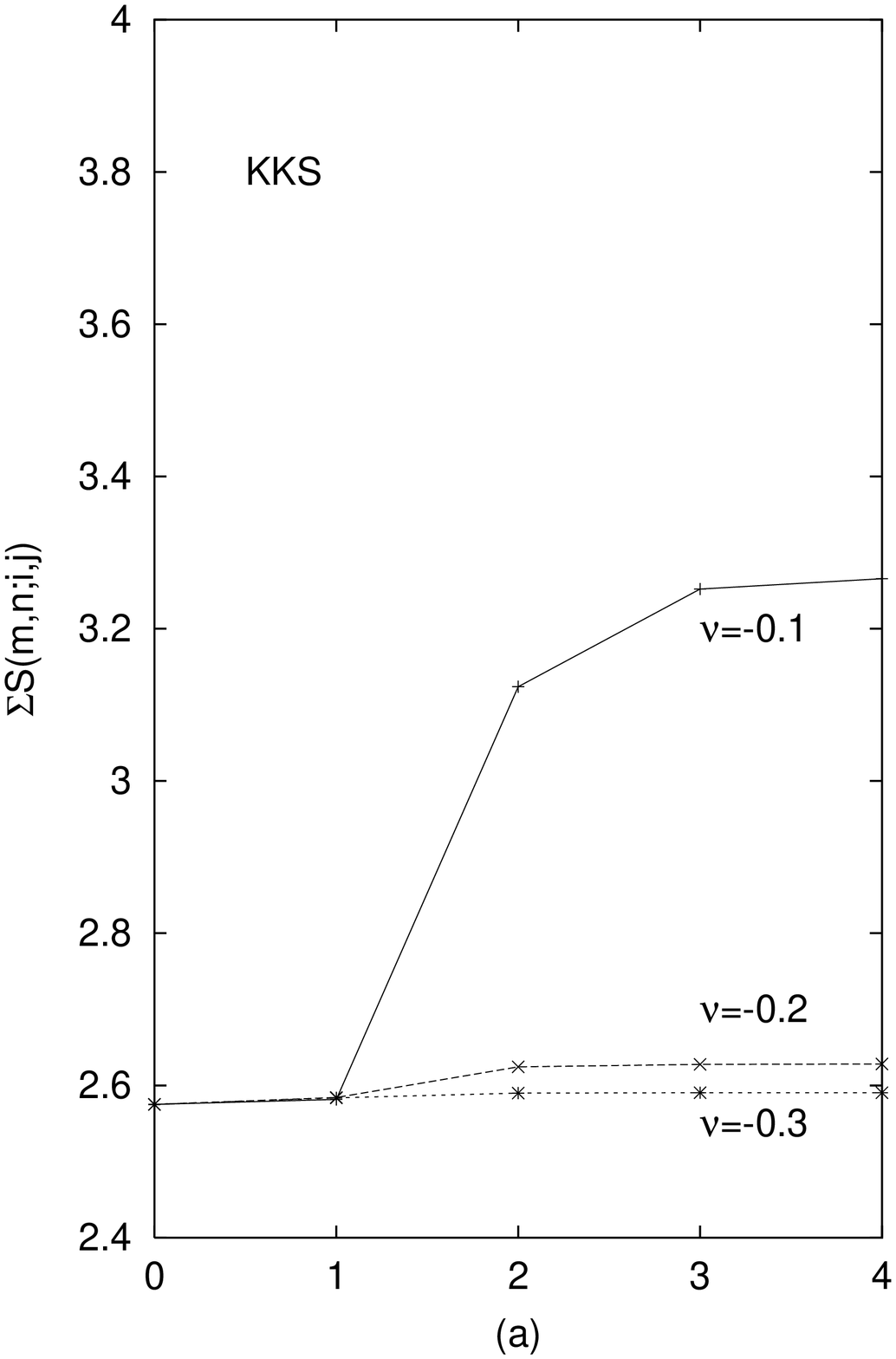,height=13cm}&\epsfig{file=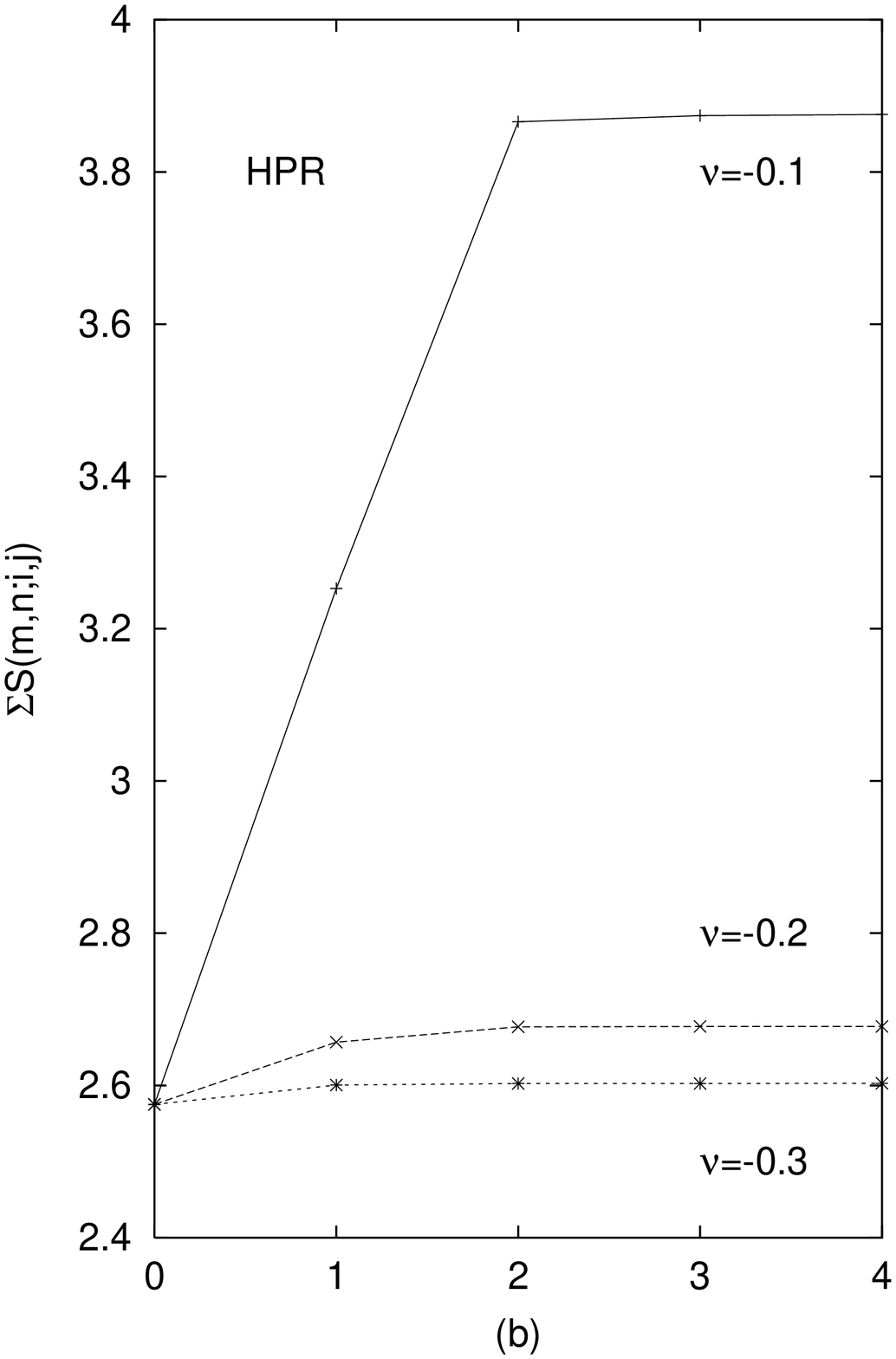,height=13cm}
\end{tabular}
\end{center}
\caption{}
\label{KKSvsDHR}
\end{figure}
\begin{figure}
\vskip 2cm
\begin{center}
\epsfig{file=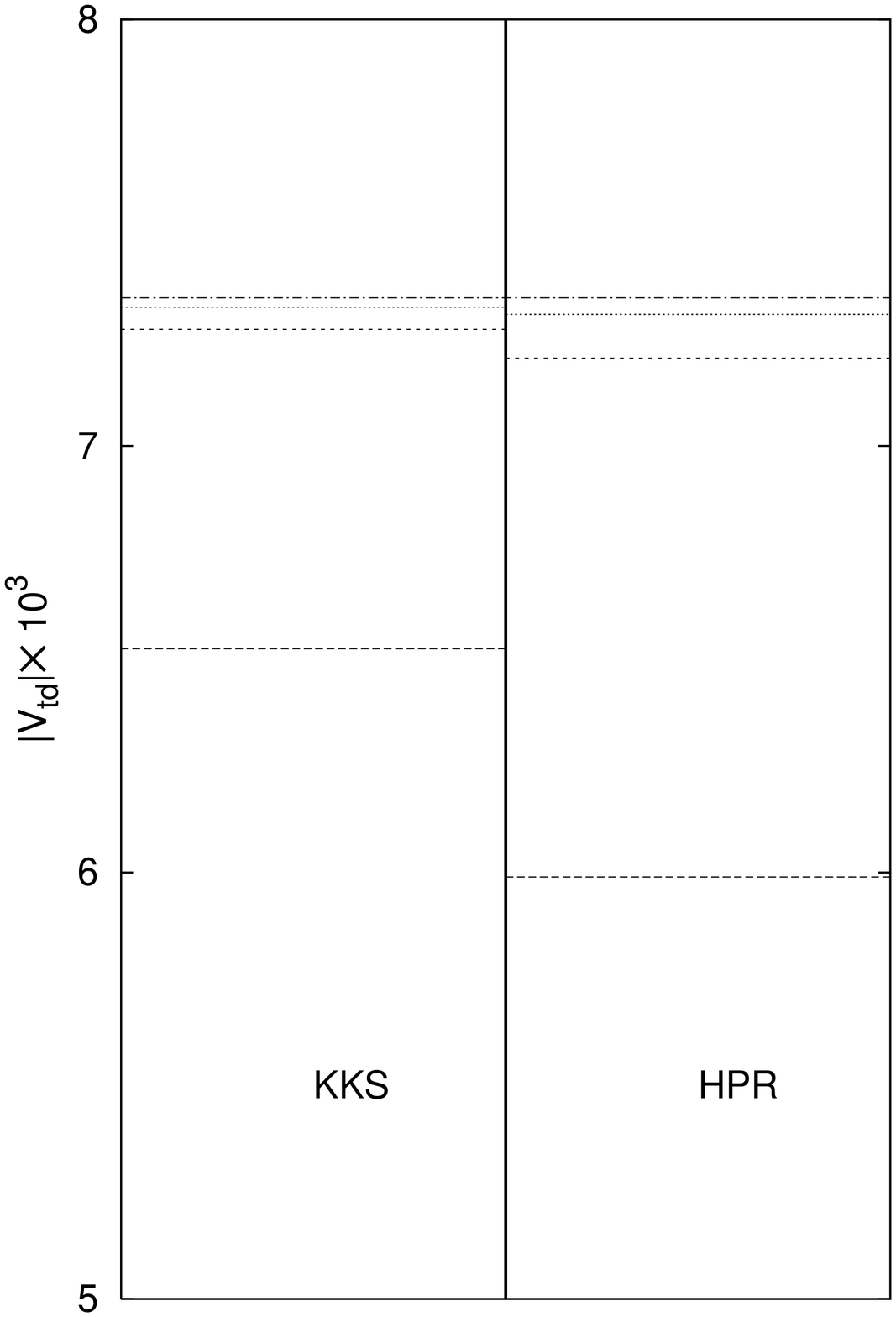,height=20cm}
\end{center}
\caption{}
\label{Vtd}
\end{figure}

In this analysis, we include the KK modes only up to $n=4$ gauge boson (which
corresponds to $n=2$ fermion because bulk fermion has double field contents) 
for simplicity.
As a comparison, the HPR contribution to the box diagrams is also given.
Numerical results are summarized in Table \ref{S-enhance} and 
Figs.\ \ref{KKSvsDHR}, \ref{Vtd}.
Main results of the analysis are as follows.
\par
First, we have the loop function enhancement $0.51\%$ ($1.07\%$) for $\nu=-0.3$
in KKS (HPR) up to $n=4$, as shown in Table 
\ref{S-enhance} and Fig.\ \ref{KKSvsDHR}.
The numbers can be compared to the ACD-UED result where the increase amounts to
17 \% \cite{Buras}.
Increase of the loop function implies the smaller value of $|V_{td}|$, because
$\Delta M_B=0.502\pm0.006~{\rm ps}^{-1}\sim |V_{td}|^2 \sum S(m,n;i,j)$ is now 
well measured \cite{HFAG}.
Figure \ref{Vtd} shows our results for $|V_{td}|$.
We used $\sqrt{{\hat B}_B}f_B=235$ MeV, $\eta_B=0.55$.
The QCD correction factor $\eta_B$ has a small error $\pm 0.01$ compared to
the hadronic uncertainty \cite{Buras,BfB}:
\begin{equation}
\sqrt{{\hat B}_B} f_B=(235^{+33}_{-41})~{\rm MeV}~.
\end{equation}
\par
Since the $B_s\Bbar_s$ mixing shares the same loop function with $B\Bbar$ 
mixing, increase in $\sum S(m,n;i,j)$ predicts the equal amount of larger
$\Delta M_{B_s}$.
Current bound of it is $\Delta M_{B_s}>15~{\rm ps}^{-1}$ \cite{HFAG}.
\par
In order to strongly constrain the model parameters, or even rule out, 
measurements of $\sin2\beta$, e.g., would not be crucial.
Rather, determining the CKM angle $\gamma$ as well as reducing the hadronic
uncertainty significantly will severely restrict the parameter space.
For example,
in the ACD-UED model, Ref.\ \cite{Chakraverty} argued that larger value of
$\sqrt{{\hat B}_B}f_B$ above 222 MeV with one-third reduced error will rule 
out the the UED model or push up the compactification scale up to multi-TeV 
region.
An analytic study is much easier in the ACD-UED model compared to the RS-type
one because the mass spectrum and mode functions have simpler structure in the
UED model \cite{ACD}.
In the RS-bulk models, on the other hand, the $n$-dependence of the mass 
spectrums, e.g., is not explicit.
\par
Second, the HPR gives larger values than KKS for each $n$.
The reason is as follows.
If all the fermions were in the bulk and no mixing happened, there would be
GIM cancellation at every $n$-th excitations.
In case of KKS, the cancellation is incomplete because there is a small mixing
(when $\nu\gtrsim -0.3$) in the top sector.
In HPR, however, the third generation resides in the 3-brane; only the first
and second generations have their KK excitations in the bulk.
There is no compensating contribution from the third generation KK modes, which
results in larger loop function.
This feature was already pointed out in \cite{KKS}.
\par
The GIM cancellation is advantageous in viewpoint of the convergence.
There is, of course, no way to warrant the finiteness of adding up infinite
KK tower by checking just a few excitations.
In the ACD-UED, it is argued that the GIM mechanism improves the convergence of 
the loop function in $B\Bbar$ mixing \cite{Buras}.
And it is quite encouraging that the RS-bulk SM effects of KKS on $
b\to s\gamma$ are satisfactorily converging as we add higher KK modes 
\cite{KKS}.
\par
Third, the $\nu$-dependence is shown in Fig.\ \ref{KKSvsDHR} and 
Table \ref{S-enhance}.
The KK mode contributions to the box diagrams get larger as $\nu$ grows.
The reason is that
both the coupling and the bulk fermion mass become larger when $\nu$ goes from 
$-3$ to $-1$.
This pattern is already studied in previous works; 
see Fig.\ 1 of \cite{DHR} and \cite{HPR}.
We have, for example, the bulk fermion masses 
$x^{(1)}_f k_{EW}=2.71,~2.85,~3.00$ (TeV) for $\nu=-0.3,~-0.2,~-0.1$, 
respectively.
It is already suggested that the plausible range of $\nu$ is 
$-0.8\lesssim\nu\lesssim-0.3$ in the literature \cite{DHR,HPR}.
If $\nu\gtrsim -0.3$, constraints from the electroweak precision data make the
first KK mode of gauge boson heavier.
In our case, since the loop function grows abruptly for large $\nu$,
present analysis of $B\Bbar$ mixing can provide another hint for the upper 
bound of $\nu$ when the uncertainties of the 
parameters for $\Delta M_B$ are reduced, just as in \cite{Chakraverty}.
\par
One thing to be noticed is that $B\Bbar$ mixing involves only the up-type 
quark's mass parameter $\nu$.
The main motivations of KKS are i) all the fermions can reside in the bulk, 
ii) different mass parameters for the top- and bottom-sector can accommodate the
electroweak precision data.
In this point of view, $D\Dbar$ is a good testing ground for the bottom sector.
A preliminary result of our approach to $D\Dbar$ mixing is that the contribution
of the first few KK modes is extremely small.
This is mainly due to the smallness of the $b$ quark mass.
Consequently, mixing with the KK masses is quite weak and the GIM cancellation 
is more complete.
Unfortunately, current experimental results are rather poor \cite{DDbar}.
More reliable check on the model can be implemented by simultaneous analysis on
$B\Bbar$ and $D\Dbar$ with improved measurements, as well as on the electroweak
precision tests.

\section{Summary}

In this paper, we have calculated the $B\Bbar$ mixing in the RS-bulk model.
Contributions from the newly introduced KK modes are positive to increase the
loop function.
We found that KKS predicts smaller contribution than HPR due to the remnant of 
the GIM cancellation.
More precise determination of the CKM unitarity triangle will test the validity
of the model.
Also, further developments of the present work, such as including much more KK
modes or simultaneous application to various observables, remain challenging.

\begin{center}
{\large\bf Acknowledgements}\\[10mm]
\end{center}

This work was supported by the BK21 Program of the Korean Ministry of Education.


\newpage

\begin{center}{\large\bf FIGURE CAPTIONS}\end{center}

\noindent
Fig.~1
\\
Bulk gauge bosons and bulk fermions contribute to the box diagrams.
\vskip .3cm
\par

\noindent
Fig.~2
\\
Enhancements of the loop function $\sum S(m,n;i,j)$ for (a) KKS and (b) HPR for
various $\nu$.
The $x$-axis is the number of KK modes included.
\vskip .3cm
\par

\noindent
Fig.~3
\\
Results for $|V_{td}|$ from the loop functions for KKS and HPR.
Each lines corresponds to $\nu=-0.1$, $-0.2$, $-0.3$ from bottom to top,
respectively.
The top line is the SM value.
\vskip .3cm
\par

\pagebreak

\begin{center}{\large\bf TABLE CAPTIONS}\end{center}
\noindent
Table 1
\\
Enhancements of the loop fuction $\sum S(m,n;i,j)$ in \% for KKS and HPR for
different values of $\nu$.
\vskip .3cm
\par\noindent



\pagebreak


\pagebreak


\begin{table}
\caption{}
\begin{tabular}{c|ccc}
 & $\nu=-0.3$ & $\nu=-0.2$ & $\nu=-0.1$ \\\hline
KKS & 0.51 & 2.06 & 26.8 \\
HPR & 1.07 & 3.98 & 50.5
\end{tabular}
\label{S-enhance}
\end{table}

%

\end{document}